\documentclass[preprint]{emulateapj}
\usepackage{hyperref}
\usepackage[usenames,dvipsnames]{color}
\hypersetup{colorlinks,citecolor=Blue,linkcolor=Red,urlcolor=Blue}
\usepackage[titletoc]{appendix}
\usepackage{amsmath}
\usepackage{float}
\usepackage{threeparttable}
\usepackage{adjustbox}

\usepackage{natbib}
\newcommand{\be}{\begin{equation}}
\newcommand{\ee}{\end{equation}}
\def\lta{\,\raise 0.3 ex\hbox{$ < $}\kern -0.75 em
 \lower 0.7 ex\hbox{$\sim$}\,}
\def\gta{\,\raise 0.3 ex\hbox{$ > $}\kern -0.75 em
 \lower 0.7 ex\hbox{$\sim$}\,} 
\newcommand{\mos}{\,m\,s$^{-1}$}
\newcommand{\kms}{\,km\,s$^{-1}$}

\newcommand\msini{\ifmmode{{\mathrm M} \sin i}\else${{\mathrm M} \sin i}$\fi}

\begin{document} 

\title{Stellar Spin-Orbit Alignment for Kepler-9, a Multi-transiting Planetary system with Two Outer Planets Near 2:1 Resonance} 

\shortauthors{Wang et al.~2017}
\shorttitle{RM effect of Kepler-9}

\author{Songhu Wang$^{1,5}$, Brett Addison$^2$, Debra A. Fischer$^1$, John M. Brewer$^1$, Howard Isaacson$^3$, Andrew W. Howard$^4$, Gregory Laughlin$^1$}
\affil{$^{1}$Department of Astronomy, Yale University, New Haven, CT 06511}
\affil{$^{2}$Department of Physics \& Astronomy, Mississippi State University, Hilbun Hall, Starkville, MS 39762}
\affil{$^{3}$Department of Astronomy, University of California, Berkeley, CA 94720}
\affil{$^{4}$Department of Astronomy, California Institute of Technology, Pasadena, CA 91125}
\affil{$^{5}$ \textit{51 Pegasi b} Fellow} 

\email{song-hu.wang@yale.edu}

\begin{abstract}

We present spectroscopic measurements of the Rossiter-McLaughlin effect for the planet b of Kepler-9 multi-transiting planet system. The resulting sky-projected spin-orbit angle is $\lambda=-13^{\circ} \pm 16^{\circ}$, which favors an aligned system and strongly disfavors highly misaligned, polar, and retrograde orbits. Including Kepler-9, there are now a total of 4 Rossiter-McLaughlin effect measurements for multiplanet systems, all of which are consistent with spin-orbit alignment.

%There are now a total of 8 spin-orbit angle measurements for multiplanet systems, including Kepler-9. The Kepler-56 system, with its stellar effective temperature of $4931{\rm K}$, remains the only unambiguous detection of a clearly misaligned multi-planetary system.

%The Kepler-56, \textbf{KELT-6?}, and \textbf{WASP-94Ab?} systems, with their host star effective temperatures of $4931{\rm K}$, $6102{\rm K}$, %and $6170{\rm K}$, remain the only unambiguous detections of clearly misaligned multi-planetary systems.
\end{abstract}

\section{Introduction}

Hot Jupiters are frequently observed to have orbital angular momentum vectors that are strikingly misaligned with their stellar spin vectors.
Stellar spin -- planetary orbit misalignments are most frequently determined through spectroscopic measurements of the Rossiter-McLaughlin (R-M) effect \citep{Rossiter1924, McLaughlin1924} during the planetary transit \citep{Queloz2000}, and have been recently reviewed by \citet{Winn2015}.

Despite years of inquiry, the origin of the spin-orbit misalignments is still unclear. Dynamically active migration mechanisms (Notably, planet-planet scattering, \citealt{Ford2008, Nagasawa2008}; Lidov-Kozai Cycling with Tidal Friction, \citealt{Wu2003, Fabrycky2007, Naoz2011}; and secular chaos, \citealt{Wu2011}), which violently deliver giant planets to short-period orbits, can naturally leave systems misaligned. In the framework of this hypothesis, the spin-orbital misalignments should represent a phenonmenon that is largely restricted to dynamically isolated hot Jupiters. 

The possibility exists, however, that the spin-orbital misalignments can be excited via mechanisms that are unrelated to planet migration. These include chaotic star formation \citep{Bate2010, Thies2011, Fielding2015} and evolution \citep{Rogers2012}, magnetic torques from host stars \citep{Lai2011}, and gravitational torques from distant companions \citep{Tremaine1991, Batygin2011, Storch2014}. In these scenarios, spin-orbit misalignments are expected to be observed not only among star-hot Jupiter pairs, but also among a broader class of planetary systems, notably those that have never experienced chaotic migration processes. This group is expected to include multiplanet systems, and especially multiplanet systems in mean motion resonance, MMR. 

\begin{deluxetable}{ccc}
\tabletypesize{\scriptsize}
\tablecaption{Radial velocity observations\label{tab1}}
\tablewidth{0pt}

\tablehead{
\colhead{Time [BJD]} & \colhead{Radial velocity [m/s]} & \colhead{Uncertainty [m/s]} 
}

\startdata
  2457959.810531  &  -3.00  &   4.93   \\
  2457959.825334  &  -5.54  &   5.15   \\
  2457959.839454  &   1.73  &   5.42   \\
  2457959.854084  &  13.12  &   5.60   \\
  2457959.868308  &   7.59  &   5.38   \\
  2457959.882834  &  -0.46  &   5.41   \\
  2457959.897139  &  21.06  &   4.67   \\
  2457959.911780  &  16.11  &   4.91   \\
  2457959.925715  &  23.18  &   4.70   \\
  2457959.940403  &  12.51  &   4.82   \\
  2457959.954766  &   7.50  &   4.58   \\
  2457959.968805  &   6.85  &   5.13   \\
  2457959.983041  & -18.62  &   5.26   \\
  2457959.997405  &  -5.29  &   5.91   \\
  2457960.012115  & -15.42  &   6.23   \\
  2457960.027092  & -15.84  &   5.49   \\
  2457960.041143  &  -0.60  &   6.00   \\
  2457960.054904  & -27.91  &   6.89   \\
  2457960.069117  &   0.10  &   8.19   \\
  2457960.084291  & -29.22  &   8.59   \\
  2457960.098284  &   8.80  &   7.74
 \enddata
\end{deluxetable} 

The R-M effect is much more easily measured when transits are frequent and deep. Therefore, as a practical consequence, although R-M observations of multiplanet systems play significant role in understanding planetary formation history, they are hard to make. They usually involve fainter stars, smaller transit depths, and/or less frequent transits, not to mention the scarcity of multiplanet systems in MMR. Although new methods (the $V\,{\rm sin}\,i$ method, \citealt{Schlaufman2010, Walkowicz2013, Hirano2014, Morton2014, Winn2017}; 
the starspot-crossing method, \citealt{Mazeh2015a, Sanchis2011, Sanchis2012, Desert2011, Dai2017}, the starspot-variability method, \citealt{Mazeh2015b}; 
the gravity-darkening method, \citealt{Barnes2009, Barnes2011, Szabo2011, Zhou2013}; 
the asteroseismic method, \citealt{Gizon2003, Chaplin2013, Van2014, Huber2013, Benomar2014}) have been developed to constrain the spin-orbit angles of multiplanet systems, as of this writing, only three robust R-M measurements exist (Kepler-89d, \citealt{Hirano2012, Albrecht2013}; WASP-47b, \citealt{Sanchis2015}; Kepler-25c, \citealt{Albrecht2013}).

In this light, Kepler-9 is particularly interesting. Kepler-9 was the first multiplanet system discovered using the transit method \citep{Holman2010}. It was also the first transiting system detected near 2:1 orbital mean motion resonance, which is believed to be the natural consequence of an evolutionary history that incorporates quiescent migration \citep{Kley2012}. Whether this system has low spin-orbit angle or not, may provide a key zeroth-order test of origin scenarios of spin-orbit misalignments and competing migration paradigms for hot Jupiters. Kepler-9~b has very large planet-star size ratio of $R_{b}/R_{*}=0.0842\pm0.0069$ \citep{Twicken2016} -- among the largest ratios yet detected in multiplanet systems. It thus offers a rare opportunity to carry out a spin-orbit angle measurement in a multiplanet system.

In this paper, we present a spin-orbit angle determination for the Kepler-9 multiplanet system that was obtained with spectroscopic R-M measurements. Our work provides additional empirical data that will further elucidate the origins of the spin-orbit misalignment distribution, and by extension, will shed light on the processes of planetary formation and evolution.

\section{Observations and Data Reduction}

\begin{figure}
\centering
\includegraphics[trim=0cm 0cm 0cm 0cm, clip=true,width=1\columnwidth]{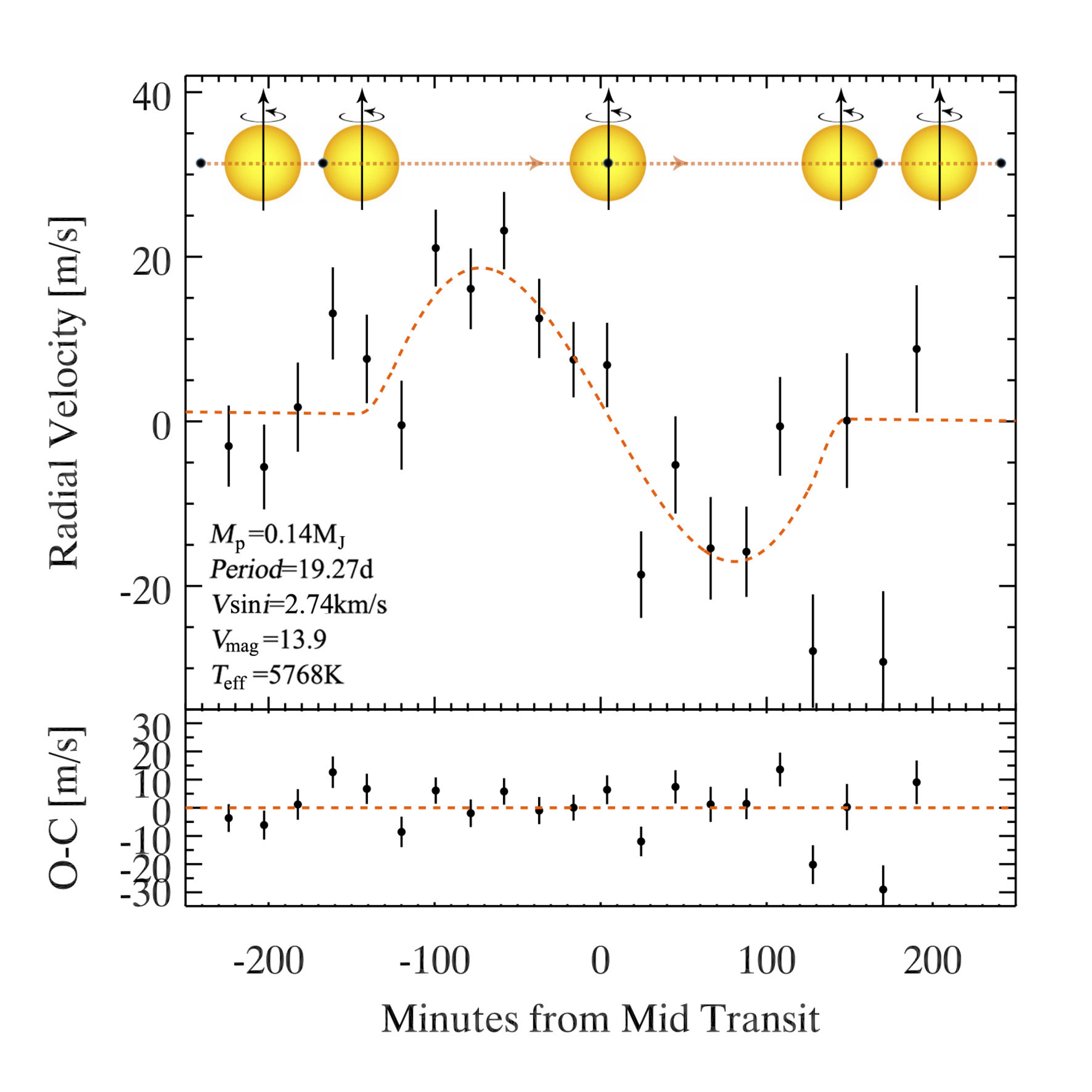}
\caption{Rossiter-McLaughlin effect observed in the Kepler-9 multiplanet system. Spectroscopic velocities spanning the transit of Kepler-9~b on the night of UT 2017 July 25 are plotted as dots with $1\,\sigma$ error bars. The orange dashed line correspond to the best-fit model which gives $\lambda=-13^{\circ} \pm 16^{\circ}$. The residuals with respect to the best-fit model are plotted in the lower panel.
}
\label{fig1}
\end{figure}

In order to measure the R-M effect, we observed the Kepler-9b transit predicted by \citet{Wang2017} to occur on the night of UT 2017 July 25 using the High Resolution Spectrograph (HIRES; \citealt{Vogt1994}) on the Keck I $10\,{\rm m}$ Telescope atop Mauna Kea in Hawaii. Although the weather was generally clear, the seeing gradually degraded over the course of the night from 0\farcs 9 to 2\farcs0.  Observations were started $1.7\,{\rm hrs}$ before the predicted time of ingress, and finished $1.1\,{\rm hrs}$ after egress (when the star set below the pointing limitation of telescope). A fraction of light is picked off behind the slit and sent to an exposure meter that individual 20-minute exposures yielded a SNR between $29-45\,{\rm pixel^{-1}}$ at $5500{\rm \AA}$. 

We obtained 21 spectra using a 0\farcs 86 slit set by the B5 decker, which provides a spectral resolution, $R\sim 55,000$. The spectra were extracted with the reduction package of the California Planet Search team \citep{Howard2010}. For each of our observations, light from the star passes through an iodine cell positioned in front of the slit. This imprints a dense forest of ${\rm I_2}$ absorption lines that are used to model the wavelength and the spectral line spread function (SLSF) of the instrument. Spectroscopic Doppler shifts were modeled using the algorithm of \citet{Butler1996} and \citet{Marcy1992}. The Doppler analysis technique uses a template spectrum of the star obtained without the iodine cell and an extremely high-resolution, high SNR Fourier Transform Spectrograph (FTS) iodine spectrum to model the observations. The best fit model is driven by a Levenburg-Marquardt least squares algorithm and is a product of the template spectrum and the FTS ${\rm I_2}$ spectrum that is then convolved with a description \citet{Valenti1995} of the SLSF. The free parameters in the model include the wavelength zero point, the dispersion, the Doppler shift and a multi-Gaussian fit to the line broadening function. At the SNR of our Kepler-9 observations, the Doppler shift was modeled with a precision of about $6\,{\rm m\,s^{-1}}$. The resulting RVs and their uncertainties are presented in Table~\ref{tab1}, and shown in Figure~\ref{fig1}.

\begin{figure}
\centering
\includegraphics[trim=0cm 0cm 0cm 0cm, clip=true,width=1\columnwidth]{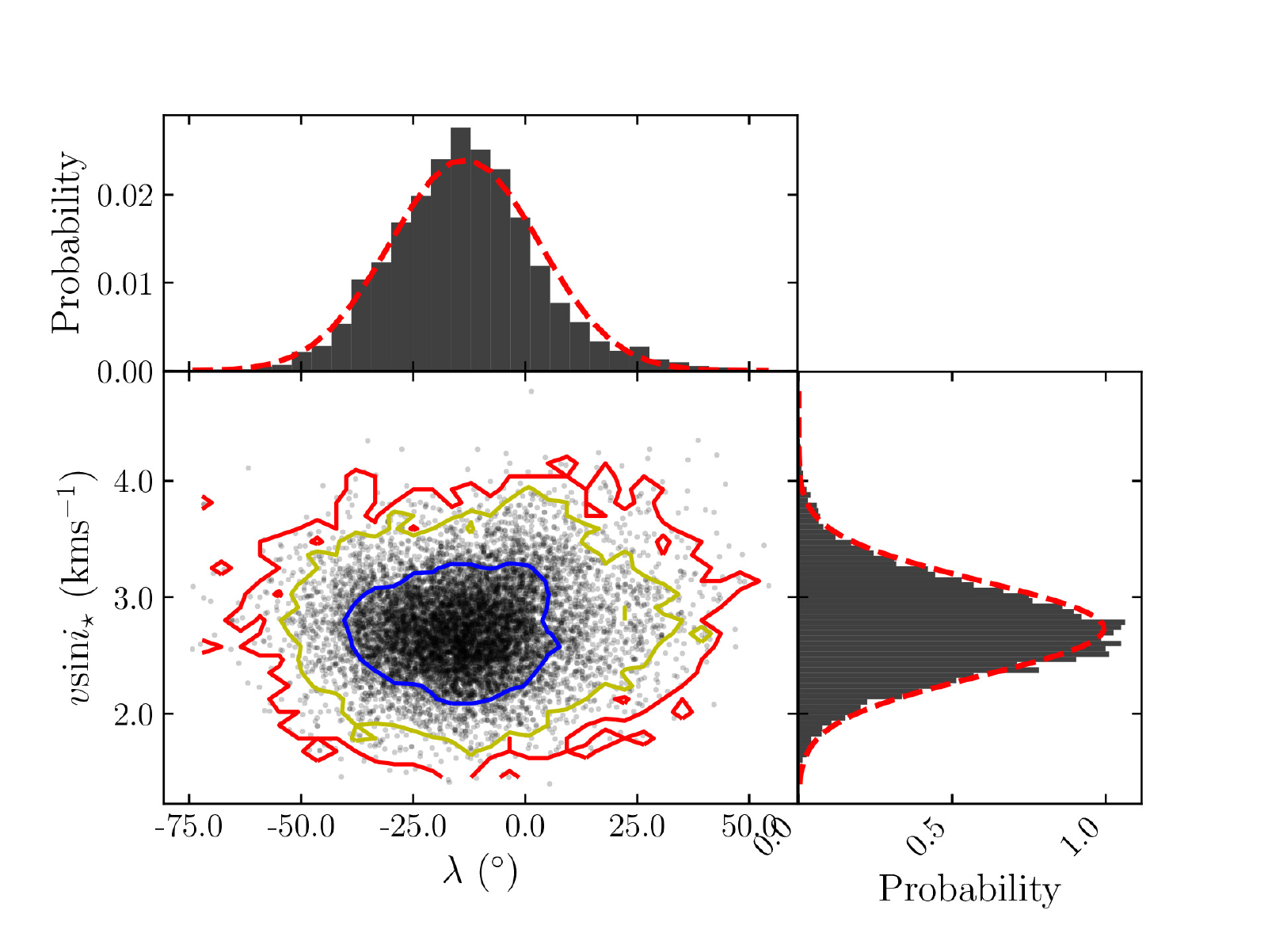}
\caption{Posterior probability distribution of $\lambda$ and $v\sin i_{\star}$ from the MCMC simulation of the Kepler-9 observations. The contours show the 1, 2, and 3 $\sigma$ confidence regions (in blue, yellow, and red, respectively). We have marginalized over $\lambda$ and $v\sin i_{\star}$ and have fit them with Gaussians (in red). This plot indicates that the distribution is Gaussian, which suggests that $\lambda$ and $v\sin i_{\star}$ are not strongly correlated with each other.}
\label{fig2}
\end{figure}

\begin{figure*}
\vspace{0cm}\hspace{0cm}
\includegraphics[scale=0.4]{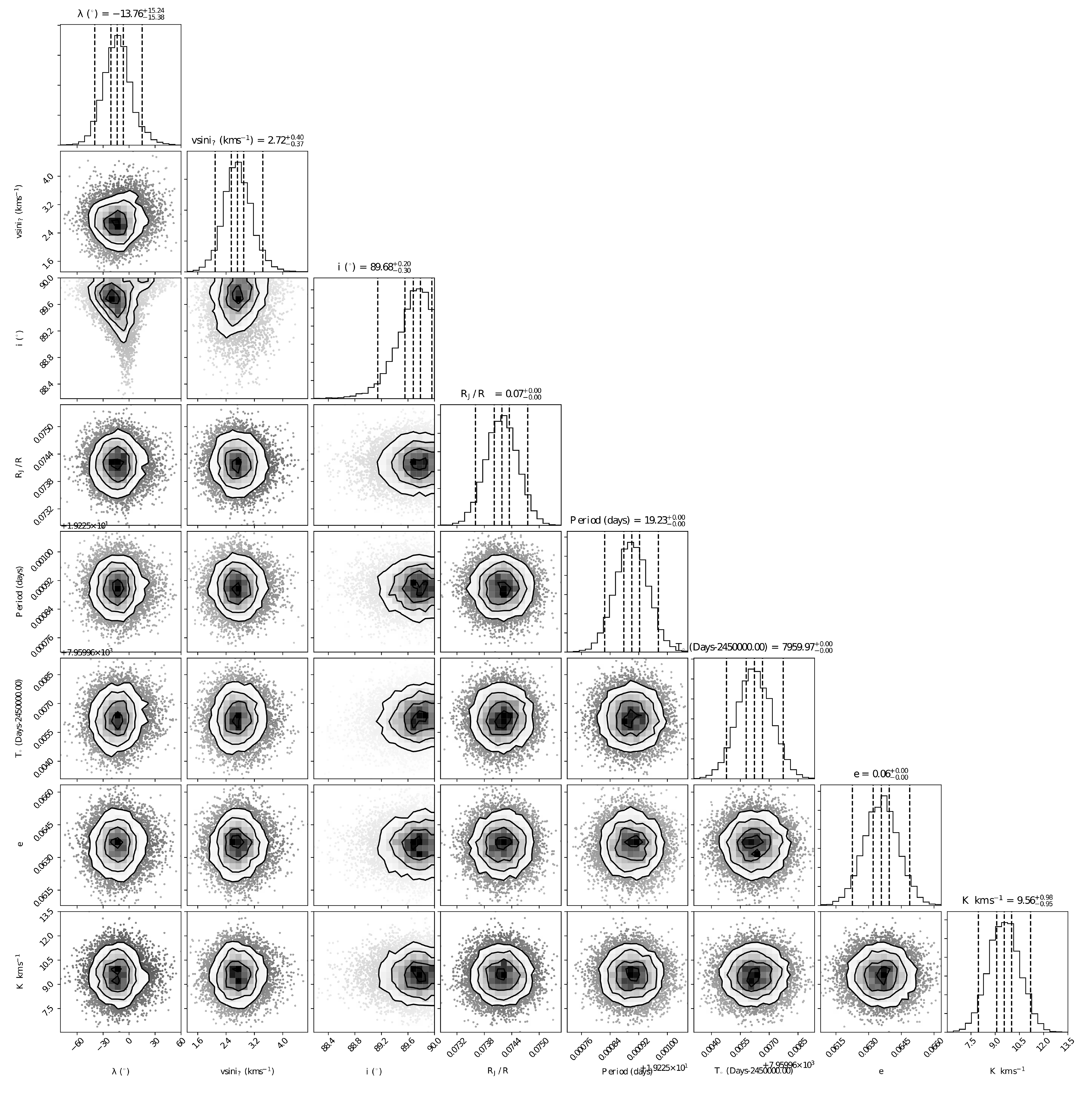}
\caption{Corner posterior probability distribution plots of the MCMC fitting parameters for the R-M anomaly fit to the Kepler-9 radial velocity data. The histograms along the diagonal show the marginalized posterior distribution for each fitting parameter. The $1\sigma$ and $2\sigma$ credibility intervals are marked by vertical dashed lines.
\label{fig3}}
\end{figure*}

\begin{figure}
\centering
\includegraphics[trim=0cm 0cm 0cm 0cm, clip=true,width=1\columnwidth]{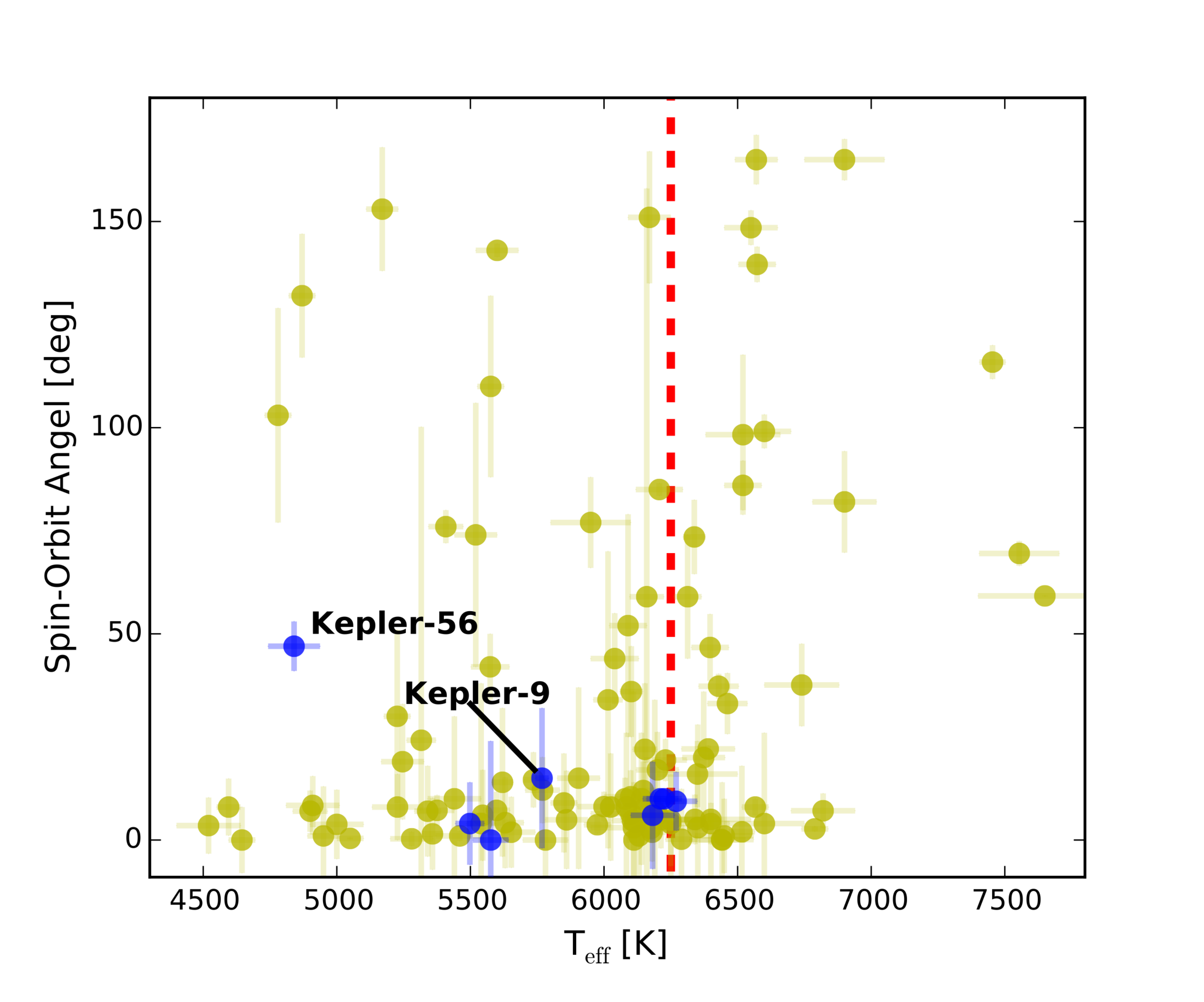}
\caption{Spin-orbit angle as a function of stellar effective temperature. Yellow dots are single-planet systems. Blue dots are multi-planet systems. The red dashed line at 6250K indicates the location of the Kraft Break \citep{Kraft1967}.}  
\label{fig4}
\end{figure}

%\begin{figure}
%\centering
%\includegraphics[trim=0cm 0cm 0cm 0cm, clip=true,width=1\columnwidth]{rm_f4.pdf}
%\caption{Posterior distribution}
%\label{fig2}
%\end{figure}

%\begin{figure}
%\centering
%\includegraphics[trim=0cm 0cm 0cm 0cm, clip=true,width=1\columnwidth]{rm_f5.pdf}
%\caption{Posterior distribution}
%\label{fig2}
%\end{figure}

\section{Analysis of the Observations}

\subsection{Independent Determination of the Projected Stellar Rotational Velocity}
We analyzed the iodine-free template observations to determine the stellar parameters and abundances using the spectral fitting procedure and line list of \citet{Brewer2016}.  The procedure has been shown to retrieve gravities consistent with those from asteroseismology to within 0.05~dex \citep{Brewer2015} in addition to accurate temperatures, precise abundances for a range of elements, and projected rotational velocities ($v \sin i$).

We first fit for the global stellar parameters, including effective temperature ($T_{\mathrm{eff}}$), surface gravity ($\log g$), metallicity ([Fe/H]), macroturbulence ($v_{mac}$), and the abundances of the alpha elements calcium, silicon, and titanium.  The initial guess for $T_{\mathrm{eff}}$ is set using the B-V color and the remaining parameters are set to solar values except for $v \sin i$, which is set to zero.  We perturb the temperature by $\pm 100$~K and re-fit, using the ${\chi}^2$-weighted average of the three fits for the input to our next step.  We then fix the global parameters, and solve for the abundances of 15 elements (C, N, O, Na, Mg, Al, Si, Ca, Ti, V, Cr, Mn, Fe, Ni, and Y).  With this new abundance pattern, we then iterate the entire procedure once.  Finally, we set the macroturbulence to using the $v_{mac}/T_{\mathrm{eff}}$ relation derived in \citet{Brewer2016} and fit for $v \sin i$. The combined uncertainties in macroturbulence and projected rotational velocity are 0.7 km/s.  Assuming equal contributions from both $v_{mac}$ and $v \sin i$ gives uncertainties of 0.5~km/s for each. Our extensive line list and differential solar analysis leads to very low statistical uncertainties in our abundances. However, model simplifications and uncertainties in the solar abundances lead to additional uncertainty in the accuracy of our abundances.  We add 0.03 dex in quadrature to the abundance uncertainties to account for the accuracy when comparing to other studies.

Our results for $T_{\mathrm{eff}}=5768\pm25{\mathrm{K}}$, $\log g=4.50\pm0.05$, ${\mathrm{[Fe/H]}}=0.035\pm0.01$, and $v \sin i=2.96\pm0.5~{\mathrm{km\,s^{-1}}}$ are, in general, consistent with recent determinations in the literature \citep{Petigura2017,Huber2014,Buchhave2012}. There is small difference in [Fe/H] as compared to \citet{Buchhave2012} at about the two sigma level.

\subsection{Determination of the Projected Stellar Obliquity}
We used the Exoplanetary Orbital Simulation and Analysis Model \citep[ExOSAM; see][]{2013ApJ...774L...9A,2014ApJ...792..112A,2016ApJ...823...29A} to determine the best-fit $\lambda$ value for Kepler-9 from the R-M effect measurements. ExOSAM utilizes a Metropolis-Hastings Markov Chain Monte Carlo (MCMC) algorithm to derive accurate posterior probability distributions of $\lambda$ and $v\sin i_{\star}$ and to optimize their fit to the RV data, largely following the procedure described in \citet{2016ApJ...823...29A}. The optimal solutions for $\lambda$ and $v\sin i_{\star}$, as well as their $1\sigma$ uncertainties, are calculated from the mean and the standard deviation of all the accepted MCMC iterations, respectively.

\begin{figure}
\centering
\includegraphics[trim=0cm 0cm 0cm 0cm, clip=true,width=\columnwidth]{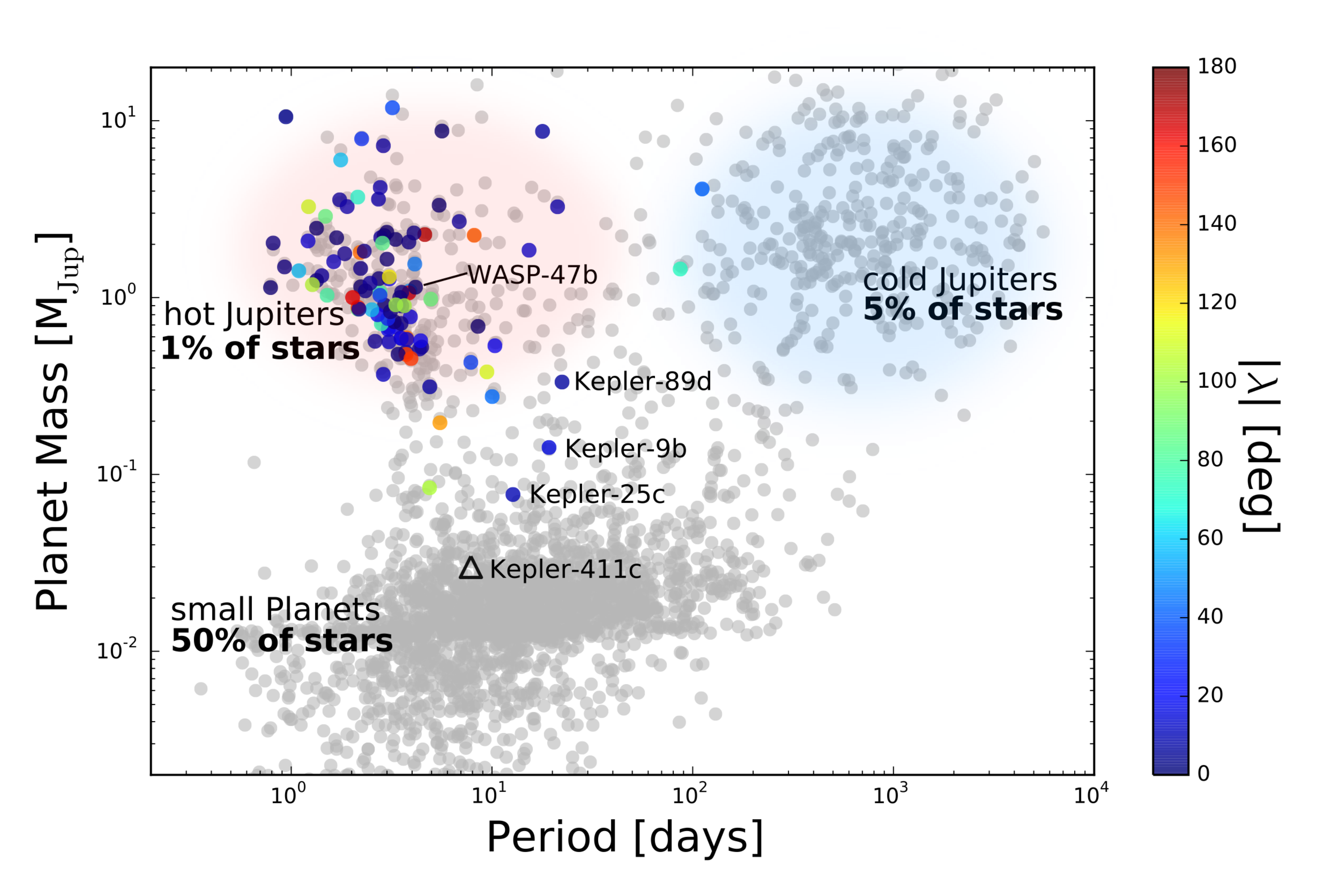}
\caption{An up-to-date mass-period diagram delineating the current extrasolar planetary catalog. Objects with Rossiter-Mclaughlin measurements (including Doppler Tomography measurements) drawn from
the TEPCat \citep{Southworth2011}\footnote{\href{http://www.astro.keele.ac.uk/jkt/tepcat/rossiter.html}{http://www.astro.keele.ac.uk/jkt/tepcat/rossiter.html}},
are shown as solid points color-coded by the measured projected spin-orbit angles, while objects without Rossiter-Mclaughlin measurements are depicted as gray transparent dots. Almost all the existing measurements are for hot Jupiters, which are intrinsically rare but readily studied. These observations have revealed a wide range of configurations of spin-orbit angles. In this work, we have expanding the list of measurements to include a member of the Kepler-9 multi-planet system, thereby probing a representative of the most populous known group of extrasolar planets.
}  
\label{fig5}
\end{figure}

Table~\ref{table:kepler-9_priors_results} lists the prior value, the $1\sigma$ uncertainty, and the prior type of each parameter used in the ExOSAM model. The results of the MCMC analysis and the best-fit values for $\lambda$ and $v\sin i_{\star}$ are also given in Table~\ref{table:kepler-9_priors_results}. For this analysis, we ran 10 independent MCMC walkers for 50,000 accepted iterations to obtain good mixing and convergence in each of the MCMC chains. 

We fixed the argument of periastron ($\omega$) given the low orbital eccentricity as well as the lack of out-of-transit RV data. We accounted for the uncertainties on $R_{\star}$ and $R_{P}$ by imposing a Gaussian prior on the planet-to-star radius ratio ($R_{P}/R_{\star}$) as well as a Gaussian prior on the stellar radius ($R_{\star}$) to account for the uncertainty in the length of the transit. Gaussian priors were imposed on the quadratic limb darkening coefficients ($q_{1}$) and ($q_{2}$) based on interpolated values from look-up tables in \citet{2011A&A...529A..75C}. We incorporated the uncertainties on the mid-transit epoch ($T_{0}$), the orbital period ($P$), orbital inclination angle ($I$), orbital eccentricity ($e$), stellar macro-turbulence ($v_\mathrm{mac}$), RV zero offset ($V_{0}$), and the stellar velocity semi amplitude ($K$) into our model using Gaussian priors from the literature. For $\lambda$, we used a uniform prior on the interval $-90^{\circ}$ to $90^{\circ}$. The Gaussian prior we placed on ${v\sin i_{\star}}$ was determined using the iodine-free spectrum template observations we obtained for Kepler-9 on the night of the transit.

Figure\,\ref{fig1} shows the modeled Rossiter--McLaughlin anomaly with the observed velocities overplotted. The Rossiter--McLaughlin effect is seen as a positive anomaly between $\sim\!150$ minutes prior to mid-transit and mid-transit and then as a negative anomaly between mid-transit and $\sim\!150$ minutes after mid-transit. This indicates that Kepler-9b first transits across the blue-shifted hemisphere during ingress and then across the red-shifted hemisphere during egress, producing a nearly symmetrical velocity anomaly. Therefore, the orbit of Kepler-9b is likely to be nearly in projected alignment with the spin axis of its host star (that is, the system is likely in ``spin--orbit alignment''). Figure\,\ref{fig1} does reveal an increase in RV scatter around egress from poorer ($\sim1.5^{\texttt{"}}$) seeing conditions.

The posterior probability distributions of $\lambda$ and $v\sin i_{\star}$, are shown in Figure \ref{fig2}. The $1\sigma$, $2\sigma$, and $3\sigma$ confidence contours are plotted, along with normalized density functions marginalized over $\lambda$ and $v\sin i_{\star}$ with fitted Gaussians. The distributions, marginalized over $\lambda$ and $v\sin i_{\star}$, adhere fairly well to a normal distribution, and appear to not be strongly correlated with each other. To check if $\lambda$ or $v\sin i_{\star}$ are strongly correlated with any of the other model parameters and to reveal covariances, we have produced a series of corner posterior probability distribution plots, which are shown in \ref{fig3}.   

\begin{table*}
\centering
\begin{adjustbox}{max width=\textwidth}
\begin{threeparttable}[b]
\caption{System Parameters, Priors, and Results for Kepler-9}
\centering
\begin{tabular}{l c c c}
\hline\hline \\ [-2.0ex]
Input Parameter & Prior & Prior Type & Results \\ [0.5ex]
\hline \\ [-2.0ex]
Mid-transit epoch (2450000-HJD), $T_{0}$ & $7959.9661 \pm 0.0009$\tnote{a} & Gaussian & $7959.9661 \pm 0.0009$ \\
Orbital period (days), $P$ & $19.225900 \pm 0.000046$\tnote{a} & Gaussian & $19.225900 \pm 0.000046$ \\
Orbital inclination, $I$ & $89.74^{\circ} \pm 0.70^{\circ}$\tnote{b} & Gaussian & $89.64^{\circ} \pm 0.26^{\circ}$ \\
Planet-to-star radius ratio, $R_{P}/R_{\star}$ & $0.074186^{+0.00022}_{-0.000348}$\tnote{b,c} & Gaussian & $0.074181 \pm 0.000351$ \\
Orbital eccentricity, $e$ & $0.0636 \pm 0.0008$\tnote{a} & Gaussian & $0.0636 \pm 0.0008$ \\
Argument of periastron, $\omega$ & $357.03^{\circ} \pm 0.44^{\circ}$\tnote{a} & Fixed\tnote{d} & -- \\
Stellar mass, $M_{\star}$ & $1.034^{+0.058}_{-0.080}$\,$M_{\odot}$\tnote{b} & Fixed & -- \\
Stellar radius, $R_{\star}$ & $0.956^{+0.147}_{-0.053}$\,$R_{\odot}$\tnote{b} & Gaussian & $ 0.982 \pm 0.068$\,$R_{\odot}$ \\
Planet mass, $M_{P}$ & $0.1384 \pm 0.0015$\,$M_{J}$\tnote{a} & Fixed\tnote{d} & -- \\
Planet radius, $R_{P}$ & $0.6905^{+0.1071}_{-0.0375}$\,$R_{J}$\tnote{b, e} & Fixed & -- \\
Impact parameter, $b$ & $0.143\pm0.022$\tnote{f} & -- & -- \\
Stellar velocity semi-amplitude, $K$ & $9.55 \pm 0.98$\,\mos\tnote{a} & Gaussian & $9.56 \pm 0.98$\,\mos \\
Stellar micro-turbulence, $\xi_{t}$ & N/A & Fixed & -- \\
Stellar macro-turbulence, $v_\mathrm{mac}$ & $3.47 \pm 0.5$\,\kms\tnote{g} & Gaussian & $3.46 \pm 0.69$\,\kms \\
Stellar limb-darkening coefficient, $q_{1}$ & $0.4699 \pm 0.0463$\tnote{h} & Gaussian & $0.4699 \pm 0.0462$ \\
Stellar limb-darkening coefficient, $q_{2}$ & $0.2507 \pm 0.0440$\tnote{h} & Gaussian & $0.2506 \pm 0.0439$ \\
RV zero offset, $V_{0}$ & $0.0 \pm 5.0$\,\mos & Gaussian & $0.5 \pm 1.4$\,\mos \\
Projected obliquity angle, $\lambda$ & $[-90^{\circ} \textendash\, 90^{\circ}]$ & Uniform & $-13^{\circ} \pm 16^{\circ}$ \\ 
Projected stellar rotation velocity, ${v\sin i_{\star}}$ & $2.96 \pm 0.50$\,\kms\tnote{g} & Gaussian & $2.74 \pm 0.40$\,\kms \\ [0.5ex]
\hline 
\end{tabular}%
%}
\label{table:kepler-9_priors_results}
\vspace{1mm}
\label{table:Kepler-9_Parameters}
\begin{tablenotes}
\item [a] \textit{Prior values determined through our dynamical simulations of transit timing variations derived from the full Kepler data set and from a photometric transit observation of Kepler-9 on UT 2016 September 1.}
\item [b] \textit{Prior values given by the NASA Exoplanet Archive in the cumulative table of planet candidates and used in the MCMC.}
\item [c] \textit{In cases where the prior uncertainty is asymmetric, for simplicity, we use a symmetric Gaussian prior with the prior width set to the larger uncertainty value in MCMC.}
\item [d] \textit{Prior fixed to allow convergence of MCMC chains.}
\item [e] \textit{Planet radius given here for informative purposes and determined from planet-to-star radius ratio prior.}
\item [f] \textit{Parameter and value given for informative purposes.}
\item [g] \textit{Priors determined from Kepler-9 spectrum template observations.}
\item [h] \textit{Limb darkening coefficients interpolated from the look-up tables in \cite{2011A&A...529A..75C}.}
\end{tablenotes}
\end{threeparttable}
\end{adjustbox}
\end{table*}

\section{Discussion}

Kepler-9 has a mass and an effective temperature that are very close to the solar values. For a number of years, as the first planetary Rossiter-McLaughlin measurements were accumulating in the literature, there was evidence that planets transiting relatively low-mass ($M\lesssim1.1M_{\odot}$), low temperature ($T\lesssim6200 K$) stars tend to have low-obliquity orbits, with the converse being true of planets orbiting higher-mass stars. Early data arguments for this picture can be found in \citet{Schlaufman2010}, \citet{Winn2010}, and \citet{Albrecht2012}.

The picture is no longer so clear-cut. The number of planet-star pairs with spin-orbit measurements has been increasing steadily, and the total number of systems with measurements is of order $N\sim120$. Figure 4 gathers the projected obliquities obtained to date, showing that while there is still an apparent statistical tendency for low-temperature stars to favor aligned orbits, the correlation has weakened substantially. As pointed out recently by \citet{Dai2017}, however, among planets with $a(1-e)\lesssim6R_{\star}$ orbiting low-mass stars, low-obliquity is still the rule. A similar pattern was also shown in \citet{Triaud2017}. This dichotomy hints at the potential importance of star-disk interactions for driving alignment in low-mass systems that had $\sim10^{3}$ gauss magnetic fields during the T-Tauri stage \citep{Dawson2014, Spalding2015}, and hints as well that star-planet tides may also be playing a coplanarizing role \citep{Winn2010, Anderson2015}.

Kepler-9b, with its long orbital period and its resonant lock to an exterior companion would likely be less prone to either evolutionary process, and one would likely retain any primordial spin-orbit misalignment. Therefore, the observed co-planarity may point to an early history in which migration and accretion occurred in isolation and with relatively little disturbance.

Finally, it is useful to note that spin-obit alignment measurements are only beginning to probe the truly representative populations of planets. As indicated by the summary diagram shown in Figure 5, the hot Jupiters (which accompany $<1$\% of stars \citep{Batalha2013}) have had their orbital obliquities sampled very heavily, but the overwhelmingly more common super-Earths and sub-Neptunes (as well as the population of longer-period Jovian planets) have as-yet barely been touched. The Kepler-9 planets lie in the sparsely populated transition region with $10\,{\rm d}\lesssim P \lesssim 100\,{\rm d}$, and $30\,{M_{\oplus}}\lesssim P \lesssim 100\,{M_{\oplus}}$. Forthcoming measurements -- such as those planned for Kepler 411-c -- will probe the great bulk of the distribution, and should clarify what happens when the planet formation process follows the apparent path of least resistance.

\begin{acknowledgments}

We are thankful to the referee Simon Albrecht for providing a thorough review that greatly improved the manuscript. We would also like to thank Dong Lai, Amaury Triaud, Christopher Spalding, Smadar Naoz, Douglas Lin, Yasunori Hori, and Hui-Gen Liu for useful discussions, as well as Yutong Wu, Xiaojia Zhang, and Dong-Hong Wu for improving the quality of the Figure.
 
 S.W. acknowledges the Heising-Simons Foundation for their generous support.

Finally, the authors wish to recognize and acknowledge the very significant cultural role and reverence that the summit of Maunakea has always had within the indigenous Hawaiian community. We are most fortunate to have the opportunity to conduct observations from this mountain.
\end{acknowledgments}


\begin{thebibliography}
%\bibliography{ms}
%A

\bibitem[Albrecht et al.(2012)]{Albrecht2012} Albrecht, S., Winn, J.~N., Johnson, J.~A., et al.\ 2012, \apj, 757, 18 

\bibitem[Albrecht et al.(2013)]{Albrecht2013} Albrecht, S., Winn, J.~N., Marcy, G.~W., et al.\ 2013, \apj, 771, 11 

\bibitem[Anderson et al.(2015)]{Anderson2015} Anderson, D.~R., Triaud, A.~H.~M.~J., Turner, O.~D., et al.\ 2015, \apjl, 800, L9 

\bibitem[Addison et al.(2013)]{2013ApJ...774L...9A} Addison, B.~C., Tinney, C.~G., Wright, D.~J., et al.\ 2013, \apjl, 774, 9

\bibitem[Addison et al.(2014)]{2014ApJ...792..112A} Addison, B.~C., Tinney, C.~G., Wright, D.~J., \& Bayliss, D.\ 2014, \apj, 792, 112

\bibitem[Addison et al.(2016)]{2016ApJ...823...29A} Addison, B.~C., Tinney, C.~G., Wright, D.~J., \& Bayliss, D.\ 2016, \apj, 823, 29

%B

\bibitem[Barnes(2009)]{Barnes2009} Barnes, J.~W.\ 2009, \apj, 705, 683 

\bibitem[Barnes et al.(2011)]{Barnes2011} Barnes, J.~W., Linscott, E., \& Shporer, A.\ 2011, \apjs, 197, 10 

\bibitem[Batalha et al.(2013)]{Batalha2013} Batalha, N.~M., Rowe, J.~F., Bryson, S.~T., et al.\ 2013, \apjs, 204, 24 

\bibitem[Bate et al.(2010)]{Bate2010} Bate, M.~R., Lodato, G., \& Pringle, J.~E.\ 2010, \mnras, 401, 1505 

\bibitem[Batygin et al.(2011)]{Batygin2011} Batygin, K., Morbidelli, A., \& Tsiganis, K.\ 2011, \aap, 533, A7 

\bibitem[Benomar et al.(2014)]{Benomar2014} Benomar, O., Masuda, K., Shibahashi, H., \& Suto, Y.\ 2014, \pasj, 66, 94

\bibitem[Brewer et al.(2015)]{Brewer2015} Brewer, J.~M., Fischer, D.~A., Basu, S., Valenti, J.~A., \& Piskunov, N.\ 2015, \apj, 805, 126

\bibitem[Brewer et al.(2016)]{Brewer2016} Brewer, J.~M., Fischer, D.~A., Valenti, J.~A., \& Piskunov, N.\ 2016, \apjs, 225, 32

\bibitem[Buchhave et al.(2012)]{Buchhave2012} Buchhave, L.~A., Latham, D.~W., Johansen, A., et al.\ 2012, \nat, 486, 375

\bibitem[Butler et al.(1996)]{Butler1996} Butler, R.~P., Marcy, G.~W., Williams, E., et al.\ 1996, \pasp, 108, 500

%C

\bibitem[Campante et al.(2016)]{Campante2016} Campante, T.~L., Lund, M.~N., Kuszlewicz, J.~S., et al.\ 2016, \apj, 819, 85 

\bibitem[Chaplin et al.(2013)]{Chaplin2013} Chaplin, W.~J., Sanchis-Ojeda, R., Campante, T.~L., et al.\ 2013, \apj, 766, 101 

\bibitem[Claret \& Bloemen (2011)]{2011A&A...529A..75C} Claret, A. \& Bloemen, S.\ 2011, \aap, 529, 75

%D

\bibitem[Dai \& Winn(2017)]{Dai2017} Dai, F., \& Winn, J.~N.\ 2017, \aj, 153, 205 


\bibitem[Dawson(2014)]{Dawson2014} Dawson, R.~I.\ 2014, \apjl, 790, L31 

\bibitem[D{\'e}sert et al.(2011)]{Desert2011} D{\'e}sert, J.-M., Charbonneau, D., Demory, B.-O., et al.\ 2011, \apjs, 197, 14 


%E

%F

\bibitem[Fabrycky \& Tremaine(2007)]{Fabrycky2007} Fabrycky, D., \& Tremaine, S.\ 2007, \apj, 669, 1298  

\bibitem[Fielding et al.(2015)]{Fielding2015} Fielding, D.~B., McKee, C.~F., Socrates, A., Cunningham, A.~J., \& Klein, R.~I.\ 2015, \mnras, 450, 3306 

\bibitem[Ford \& Rasio(2008)]{Ford2008} Ford, E.~B., \& Rasio, F.~A.\ 2008, \apj, 686, 621 

%G

\bibitem[Gizon \& Solanki(2003)]{Gizon2003} Gizon, L., \& Solanki, S.~K.\ 2003, \apj, 589, 1009 

%H

\bibitem[H{\'e}brard et al.(2008)]{Hebrard2008} H{\'e}brard, G., Bouchy, F., Pont, F., et al.\ 2008, \aap, 488, 763 

\bibitem[Hirano et al.(2012)]{Hirano2012} Hirano, T., Narita, N., Sato, B., et al.\ 2012, \apjl, 759, L36 

\bibitem[Hirano et al.(2014)]{Hirano2014} Hirano, T., Sanchis-Ojeda, R., Takeda, Y., et al.\ 2014, \apj, 783, 9

\bibitem[Holman et al.(2010)]{Holman2010} Holman, M.~J., Fabrycky, D.~C., Ragozzine, D., et al.\ 2010, Science, 330, 51 

\bibitem[Howard et al.(2010)]{Howard2010} Howard, A.~W., Johnson, J.~A., Marcy, G.~W., et al.\ 2010, \apj, 721, 1467 

\bibitem[Huber et al.(2014)]{Huber2014} Huber, D., Silva Aguirre, V., Matthews, J.~M., et al.\ 2014, \apjs, 211, 2

\bibitem[Huber et al.(2013)]{Huber2013} Huber, D., Carter, J.~A., Barbieri, M., et al.\ 2013, Science, 342, 331 

%I

%J

%K

\bibitem[Kley \& Nelson(2012)]{Kley2012} Kley, W., \& Nelson, R.~P.\ 2012, \araa, 50, 211 

\bibitem[Kraft(1967)]{Kraft1967} Kraft, R.~P.\ 1967, \apj, 150, 551 

%L

\bibitem[Lai et al.(2011)]{Lai2011} Lai, D., Foucart, F., \& Lin, D.~N.~C.\ 2011, \mnras, 412, 2790 

%M
\bibitem[Marcy \& Butler(1992)]{Marcy1992} Marcy, G.~W., \& Butler, R.~P.\ 1992, \pasp, 104, 270

\bibitem[Mazeh et al.(2015)]{Mazeh2015a} Mazeh, T., Holczer, T., \& Shporer, A.\ 2015a, \apj, 800, 142 

\bibitem[Mazeh et al.(2015)]{Mazeh2015b} Mazeh, T., Perets, H.~B., McQuillan, A., \& Goldstein, E.~S.\ 2015b, \apj, 801, 3 

\bibitem[McLaughlin(1924)]{McLaughlin1924} McLaughlin, D.~B.\ 1924, \apj, 60, 

\bibitem[Morton \& Winn(2014)]{Morton2014} Morton, T.~D., \& Winn, J.~N.\ 2014, \apj, 796, 47 

%N

\bibitem[Nagasawa et al.(2008)]{Nagasawa2008} Nagasawa, M., Ida, S., \& Bessho, T.\ 2008, \apj, 678, 498 

\bibitem[Naoz et al.(2011)]{Naoz2011} Naoz, S., Farr, W.~M., Lithwick, Y., Rasio, F.~A., \& Teyssandier, J.\ 2011, \nat, 473, 187 

%O

%P

\bibitem[Petigura et al.(2017)]{Petigura2017} Petigura, E.~A., Howard, A.~W., Marcy, G.~W., et al.\ 2017, \aj, 154, 107 

%Q

\bibitem[Queloz et al.(2000)]{Queloz2000} Queloz, D., Eggenberger, A., Mayor, M., et al.\ 2000, \aap, 359, L13 

%R

\bibitem[Rogers et al.(2012)]{Rogers2012} Rogers, T.~M., Lin, D.~N.~C., \& Lau, H.~H.~B.\ 2012, \apjl, 758, L6

\bibitem[Rossiter(1924)]{Rossiter1924} Rossiter, R.~A.\ 1924, \apj, 60,

%S

\bibitem[Sanchis-Ojeda et al.(2011)]{Sanchis2011} Sanchis-Ojeda, R., Winn, J.~N., Holman, M.~J., et al.\ 2011, \apj, 733, 127 

\bibitem[Sanchis-Ojeda et al.(2012)]{Sanchis2012} Sanchis-Ojeda, R., Fabrycky, D.~C., Winn, J.~N., et al.\ 2012, \nat, 487, 449

\bibitem[Sanchis-Ojeda et al.(2015)]{Sanchis2015} Sanchis-Ojeda, R., Winn, J.~N., Dai, F., et al.\ 2015, \apjl, 812, L11

\bibitem[Schlaufman(2010)]{Schlaufman2010} Schlaufman, K.~C.\ 2010, \apj, 719, 602 

\bibitem[Southworth(2011)]{Southworth2011} Southworth, J.\ 2011, \mnras, 417, 2166 

\bibitem[Spalding \& Batygin(2015)]{Spalding2015} Spalding, C., \& Batygin, K.\ 2015, \apj, 811, 82 

\bibitem[Storch et al.(2014)]{Storch2014} Storch, N.~I., Anderson, K.~R., \& Lai, D.\ 2014, Science, 345, 1317 

\bibitem[Szab{\'o} et al.(2011)]{Szabo2011} Szab{\'o}, G.~M., Szab{\'o}, R., Benk{\H o}, J.~M., et al.\ 2011, \apjl, 736, L4
%T



\bibitem[Thies et al.(2011)]{Thies2011} Thies, I., Kroupa, P., Goodwin, S.~P., Stamatellos, D., \& Whitworth, A.~P.\ 2011, \mnras, 417, 1817 

\bibitem[Triaud(2017)]{Triaud2017} Triaud, A.~H.~M.~J.\ 2017, arXiv:1709.06376 

\bibitem[Tremaine(1991)]{Tremaine1991} Tremaine, S.\ 1991, \icarus, 89, 85 

\bibitem[Twicken et al.(2016)]{Twicken2016} Twicken, J.~D., Jenkins, J.~M., Seader, S.~E., et al.\ 2016, \aj, 152, 158 

%U


%V

\bibitem[Valenti et al.(1995)]{Valenti1995} Valenti, J.~A., Butler, R.~P., \& Marcy, G.~W.\ 1995, \pasp, 107, 966

\bibitem[Van Eylen et al.(2014)]{Van2014} Van Eylen, V., Lund, M.~N., Silva Aguirre, V., et al.\ 2014, \apj, 782, 14 

\bibitem[Vogt et al.(1994)]{Vogt1994} Vogt, S.~S., Allen, S.~L., Bigelow, B.~C., et al.\ 1994, \procspie, 2198, 362 

%W

\bibitem[Walkowicz \& Basri(2013)]{Walkowicz2013} Walkowicz, L.~M., \& Basri, G.~S.\ 2013, \mnras, 436, 1883 

\bibitem[Wang et al.(2017)]{Wang2017} Wang, S., Wu, D.-H., Addison, B., et al.\ 2017, Submitted.

\bibitem[Winn et al.(2010)]{Winn2010} Winn, J.~N., Fabrycky, D., Albrecht, S., \& Johnson, J.~A.\ 2010, \apjl, 718, L145 

\bibitem[Winn \& Fabrycky(2015)]{Winn2015} Winn, J.~N., \& Fabrycky, D.~C.\ 2015, \araa, 53, 409

\bibitem[Winn et al.(2017)]{Winn2017} Winn, J.~N., Petigura, E.~A., Morton, T.~D., et al.\ 2017, arXiv:1710.04530 

\bibitem[Wu \& Murray(2003)]{Wu2003} Wu, Y., \& Murray, N.\ 2003, \apj, 589, 605 

\bibitem[Wu \& Lithwick(2011)]{Wu2011} Wu, Y., \& Lithwick, Y.\ 2011, \apj, 735, 109 


%X

%Y

%Z

\bibitem[Zhou \& Huang(2013)]{Zhou2013} Zhou, G., \& Huang, C.~X.\ 2013, \apjl, 776, L35

\end{thebibliography}
\end{document}